\documentclass[a]{iucr}

\usepackage{url}
\usepackage{graphicx}

\usepackage{color}

\usepackage{needspace}
\renewcommand{\ack}[1]{\needspace{6em}%
{\medskip\noindent\sffamily\normalsize\bfseries Acknowledgments}%
\smallskip\par\noindent#1%
}

\journalcode{J} 

\begin{document} 



\title{\bf\emph{reductus}: a stateless Python data-reduction service with a browser frontend}


\cauthor[a]{Brian}{Maranville}{brian.maranville@nist.gov}{}
\author[a]{William}{Ratcliff II}
\author[a]{Paul}{Kienzle}

\aff[a]{%
 NIST Center for Neutron Research
 100 Bureau Drive, Gaithersburg MD USA 20899}%








\maketitle                        

\begin{synopsis}
A web-based, flexible scientific data reduction system is presented, which accesses published data stores and transforms raw measurements into interpretable data through data-flow diagrams which are converted to advanced calculations on a Python backend; the results are returned in real time through the web interface.  The application was  developed for handling neutron and X-ray reflectometry results at a user facility.
\end{synopsis}

\begin{abstract}
The online data reduction service \emph{reductus} transforms
measurements in experimental science from
laboratory coordinates into physically meaningful quantities
with accurate estimation of uncertainties based on instrumental
settings and properties.
This reduction process is based on a few well-known
transformations, but flexibility in the application of the transforms
and algorithms supports flexibility in experiment design, supporting
a broader range of measurements than a rigid reduction scheme for data.
The user interface allows easy construction of arbitrary
pipelines from well-known data transforms using a visual dataflow diagram.
Source data is drawn from a networked, open data repository.
The Python backend uses intelligent caching to store intermediate results
of calculations for a highly responsive user experience.
The reference implementation allows immediate reduction of measurements
as they are recorded for the three neutron reflectometry instruments at the NIST
Center for Neutron Research (NCNR), without the need for visiting scientists to install
additional software on their own computers.
\end{abstract}

\section{Motivation}

The transformation of raw measurement output into meaningful, interpretable
data with attached uncertainties (data reduction)  is a ubiquitous task in the experimental sciences.
In the case where the workflow is well-established and the community is small the most direct way to accomplish this is to develop a custom application to be installed on a limited number of dedicated computers.
However, at a scientific user facility a large number of visiting researchers  are using the measurement tools on a part-time basis.
As such, it is necessary to make reduction capabilities widely available and flexible.
In this case a web-based application is an attractive alternative to distributing dedicated installable executables.

The main benefit of a web application is the almost universal accessibility to potential users.
On top of this, a centralized reduction server also benefits from the ability to update the calculation code at any time without requiring all users to update their software, as well as largely eliminating the non-trivial time cost of maintaining an installable application for more than one target platform.
Users can also install the service locally much as they would if reduction were a desktop application.
This allows them to access their own data and develop reduction applications for their own instrumentation.

The specific implementation described herein was developed to provide reduction for reflectometry instruments, but the system was designed to be extensible and usable for other data processing problems and work is underway to support reduction for off-specular reflectometry, small-angle neutron scattering (SANS), and triple axis spectrometry (TAS).
The approach can be adapted to any type of computational problem, but the level of flexibility and computational complexity supported is driven by the needs of a data reduction engine.

\subsection{Data reduction for reflectometry}

The measurements for reflectometry are performed at a particular configuration in laboratory coordinates (primarily incident and detector angles) as recorded in the raw data files, with some uncertainty in the precise values.\cite{dura2006,salah2007}
In counting experiments with a beam of particles incident on the sample there are many independent events for each configuration, each with a slightly different wavelength and flight path.
When accumulated, this variation manifests as instrument resolution, which can be convolved with the theoretical model of the experiment to produce an expected count rate for that configuration.
Usually the experiment will be set up so that configuration uncertainty is much less than instrument resolution in order to maximize the information gained from the measurement, and so configuration uncertainty is largely ignored.

In order to determine the relative portion of scattered counts for each instrument configuration, the total incident beam rate must be measured independently, then simple division can normalize to a relative count rate.
With the incident beam generated according to a Poisson distribution,
the uncertainty in counts grows as $\surd n$, which leads to count rate estimates which grow as $1/\surd n$; counting longer leads to smaller error bars.

There will be some background count rate, either from alternative paths to the detector such as scattering through air, or from completely distinct processes such as electronic noise, which can be subtracted from the measured scattering signal giving an estimate of the true signal.
Background subtraction can lead to negative values, which is an unphysical estimate of the true count rate.
A more complete analysis of the data would model both the signal and background, and determine parameter estimates consistent with both the background measurement alone and with the combined signal and background measurement.
By exposing the complete reduction chain the user is free to choose the style of analysis.


\section{Dataflow diagram as a template for computation}

\begin{figure}
\label{fig:simple_diagram}
\caption{
(a) Measured signal and background counts.
(b) Dataflow diagram, with the left boxes for each node representing the input data and the right boxes representing the possible outputs.
The experimentalist can select the signal, background and normalization as input files, producing the reduced data as the output for the \texttt{divide} node.
(c) Reduced output.
}
\includegraphics[width=\linewidth]{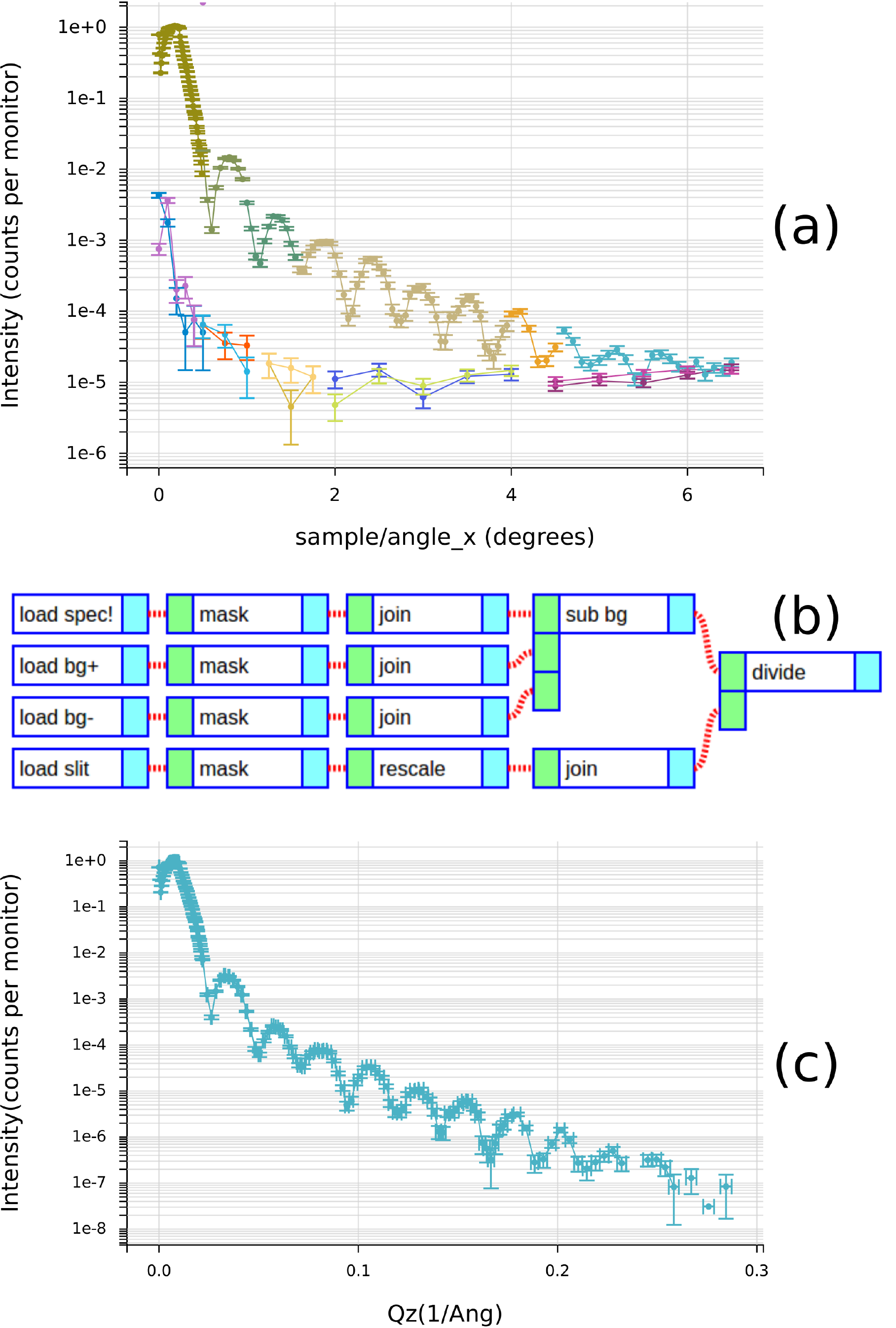}
\end{figure}

The reduction process can be described as a series of data transformations,
with data flowing from one transformation module into the next.
Corrections such as background subtraction and incident rate normalization
can be chained together in a dataflow diagram,\cite{sutherland1966}
with the final data set saved as the result of the reduction.
This yields the estimated count per incident count for each configuration.
The incident angles and wavelengths converted to more meaningful reciprocal space $q$
(inverse distance) coordinates (see Figure \ref{fig:simple_diagram}).
The individual modules may have control parameters, such as a scale factor for
a scaling module (needed if there is an attenuator in one measurement but not in the others).

\subsection{Data types}

The data flowing between modules has a type associated with it.
In order to connect the output of one module to the input of the next module,
the type of the output parameter on the first module must match the type of
the input parameter on the subsequent module.
For each data type, there are methods to store and load the data,
to convert the data to display form, and to export the data.
The stored format should be complete, so that reloading saved data
returns an equivalent data set.  The displayed and exported forms
might be a subset of the total data.

\subsection{Operations}

By implementing the transformation modules in Python, the instrument scientist has access to a
large library of numerical processing facilities based on NumPy
and SciPy,\cite{oliphant2007python} and in particular, libraries which
support propagation of uncertainties.\cite{lebigot2013uncertainties}
The \emph{reductus} library includes additional facilities for unit conversion,
data rebinning, interpolation and weighted least squares solving as well
as its own simplified uncertainties package for handling large data sets.

\subsection{Bundles of inputs} 

It is often the case that many measurements need to be combined,
with the same computation steps applied to each data file.
Rather than defining a separate chain for each file, \emph{reductus} instead
sends bundles of files between nodes.
To interpret the bundles, the module parameters are defined as either single or multiple.
If the primary input is single, then the module action
operates on each of the inputs separately; if multiple, then all inputs are
passed to the module action as a single list.
For example, if several measurements must be scaled independently then joined together into a single
data set, both the scale module input would be tagged as single,
but the join module input would be tagged multiple.
The scale factor would be tagged multiple, indicating a separate scale factor for each input.
Outputs can also be single or multiple.
Unlike the join module, which produces a single output from multiple inputs,
a split module would produce multiple outputs from a single input.
A module which rebalances data amongst inputs ({\it e.g.,} to correct for leakage between
polarization states in a polarized beam experiment), takes multiple inputs and
produces multiple outputs.

\subsection{Instrument and module definition}

An instrument is a set of data types and the computation modules
for working with them.
A computation module has a number of properties, including
name, description, version, module action and parameters.
Each parameter has an id, a label, a type, a description,
and some flags indicating whether the parameter is optional or
required, and if it is single or multiple.
Input and output parameters use one of the data types defined for the instrument.
Control parameters can have a variety of types, including
simple integers, floats or strings, or more complicated types
such as indices into the data set or coordinates on the graph,
allowing some parameter values to be set with mouse pointer
interaction in the user interface.

\subsection{Module interface language}

To encourage accurate and complete module documentation, the module
interface can be extracted from the stylized module documentation
and function definition.
The module interface language starts with an overview of the module action.
For each input, control and output parameter it gives the data type and units
and provides a short description of the parameter which can be displayed as
a tool tip in the user interface.
Inputs and control parameters are distinguished by examining the
action declaration; the positional parameters are inputs that
can be wired to another node's output and the keyword parameters
are control parameters.
After the parameters, the module documentation should define the author and version.
The module name is set to the name of the action function.

The module interface language is valid ReStructured Text (RST),
which means that the standard \texttt{docutils} toolset for Python can be
used to convert the documentation string to hypertext markup (HTML)
or portable document format (PDF).
The conversion to HTML is performed with \texttt{Sphinx}, allowing
for the creation of an independent user manual for each instrument;
it is also done dynamically for each module for display in the user interface.
Embedded equations are rendered in HTML using \texttt{mathjax},
a \TeX\ equation interpreter for javascript.

\subsection{Serialization of the diagram}

A dataflow diagram is represented as a list of nodes numbered from $0$ to $n$,
with each node having a computation module, a label, an $(x,y)$ position,
and values for the control parameters of the computation module.
The connections are defined as a list of links, with each link having
a source (node number and output parameter name),
and a target (node number and input parameter name).

Every diagram can be used as a template, with the configuration values
for the nodes packaged separately from the diagram.
The computation engine looks first in the configuration for control parameter values for the node,
using the value given in the diagram if a specific configuration value is not found.
If no value is provided in the configuration or in the diagram
then the default parameter value for the module is used.

\section{Backend}

\begin{figure}
\label{fig:architecture}
\caption{\emph{reductus} system diagram.
Upon receiving a request from the User Interface, the Load Balancer on
the Web server will find an available Python thread to run the reduction diagram.
The first step will be to fetch the requested data files from the data source and save them in the Redis Cache.
Intermediate calculations may also be cached allowing future repeated requests to be returned immediately to the client, trading efficiency against the size of the cache.
As demand increases the different parts can be run on different servers to spread the load.
}
\includegraphics[width=\linewidth]{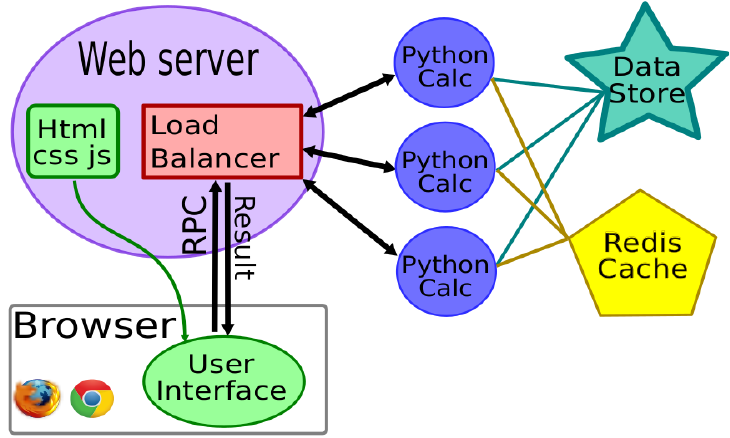}
\end{figure}

The backend is composed of two main pieces: a traditional web (HTTP) server
providing static resources including HTML, javascript,
and cascading style sheet (CSS) source code for the client application
in the user's browser, and a computation engine that
handles requests for reduction calculations as remote procedure calls (RPC), as
diagrammed in Fig. \ref{fig:architecture}.
A pool of shared calculation engines is shown but is only recommended for a production server,
as a single engine is sufficient for a single-user test environment.
A shared, disk-backed computation cache is not required but is strongly recommended for a
responsive server, even in a single-user environment
(a per-instance, non-persistent in-memory cache is the fallback option.)

The \texttt{Data Store} in Fig. \ref{fig:architecture} is not part of the server
but is an HTTP-accessible source of raw data%
\footnote {for example, the NIST Center for Neutron Research data store is located at
\url{https://dx.doi.org/10.18434/T4201B}}
which is loaded as the first step of a reduction.
This arrangement makes it possible to do data reduction without
handling the raw data files on the client computer---the user
can download just the reduced data if they wish.

\subsection{Computation server}

The point of contact for the client is the web server, which serves
the static resources (HTML, JS, CSS) as well as being a proxy
gateway to the calculation engines through
the Python web services gateway interface (WSGI).

\subsubsection{Converting the diagram to computations}
\label{sec:computing_templates}

A dataflow diagram is a directed acyclic graph (DAG), with the modules as the
nodes in the graph and the connections from outputs to inputs as the
links between the nodes.
No cycles are allowed, which means that the output of a module cannot be used to compute its own input.
Every DAG has topological order, so the nodes can be arranged linearly such that the independent nodes
appear first, and each dependent node appears after all of its input nodes.
By computing nodes in topological order all inputs are guaranteed to be
computed before they are needed.
Although there are linear time algorithms for sorting a DAG, the
diagram sizes for data reduction are small enough that a na\"ive $O(n^2)$
algorithm can be used.

\subsubsection{Results and source caching}
The results of every calculation request are cached on the server.
There are several choices in the configuration of the server
but the default is to use a \texttt{Redis}
\cite{sanfilippo2012redis} key-value store with a
Least-Recently-Used (LRU) expiry algorithm and a maximum cache size.  This can
be started automatically as a system service at startup on a Linux server,
allowing worry-free operation between server reboots.

The server cache is very important to the performance of the service: the slowest part
of many computations is retrieving the source data files from the data repository
to the server.  With caching, this typically will only happen once per data file per
experiment, and after that the server-local cached version will be used.  For data files, it is
assumed that they can be uniquely identified by their resource path (URL) and last-modified time,
(accurate to a second due to the file system \texttt{mtime} limitations).  It is therefore fallible
if the file is rewritten in a time span of less than a second, but the data files we
are using are updated much more slowly than that.

Each calculation step is identified by a unique id, created by hashing all
of the input values along with the version number of the code for the step itself.
For inputs which are the result of a previous calculation step, the hash of that
step is used to identify the input.  Since the calculations on the same data should
give the same results each time, the results can be cached using this key and
retrieved for future calculation steps or returned to the web client for display.
If the calculation parameters change (for example, the scale factor for a scaling step),
then the hash will change, and the step will be recalculated and cached with the new key.
This will change the input values for the subsequent step, which will change its hash,
which will cause it to be recalculated as well.  In this way, if the source
data (timestamp) changes, all reduction steps which depend upon it will be updated.
By including the version number of each step in the cache, changes to the server will
automatically cause all future reductions to be recomputed, even if they are already
cached.

\subsubsection{Data Provenance and Reproducibility}

In a highly interactive environment, where parameters, files and even work flows can be modified and the results saved at any time, it is important to have a record of the inputs as well as the results.\cite{simmhan2005}
Therefore, the reduction template and all its parameters are stored along with the reduced data in each saved file.
By loading a file into \emph{reductus} the precise steps required to reproduce the data become visible.
The NCNR data source is referenced through a Digital Object Identifier (DOI), with the implicit promise of a stable reference even if the data is moved to a new URL, thus providing long term reproducibility of the reduction.

The server uses the current version of each reduction step to evaluate the outputs.
This effectively acts as a behind-the-scenes update to all steps in the reduction process; any steps that are newer will be recomputed, and the updated results can be re-exported.
This is particularly useful for reductions performed during data acquisition.
As newer measurements are added the updated timestamp will force recomputation of all subsequent steps.

In order to reproduce an existing reduction, the server version at the time of the reduction must be used.
The server source is managed with the \texttt{git} source control system,\cite{torvalds2010git}
available on GitHub at \url{https://github.com/reductus/reductus}.
Git creates a unique hash of the entire source tree for each commit which is stored as part of the template.
To reproduce the data from a specific reduction this hash must be used to retrieve the source code and run it as a local server.
The specific versions of the dependencies
(\texttt{scipy}, \texttt{numpy}, \texttt{uncertainties}, $\ldots$)
can be recorded in the source tree as well to protect against changing interfaces.
Because users can easily revert to older versions of the software, developers are free to modify the code at will and maintain a clean code base.


%

\subsubsection{Statelessness}

The computation engine maintains no state.
The user interface manages the interactions of the end-user with the engine, and keeps a working copy of the
active dataflow template(s); the browser session is the only real context.
This has distinct operational advantages for the compute engine---it can be
restarted at any time with close to zero impact on the availability and continuity
of the service.  The cache is persistent
between engine restarts, but can be completely wiped if necessary without destroying
user data (performance will suffer temporarily as the calculations are re-cached).

\subsection{Server configurations}

The system is designed to be modular, allowing a number of
possible configurations of the needed resources.

\subsubsection{Simple single-computer configuration}
The simplest configuration is
to run the web server, calculation nodes and cache on the same
computer.
A server implementation using the Python \texttt{flask} package
is provided, which can simultaneously serve the static client resources
as well as the RPC calculation requests.
This server is suggested for use when running a private reduction service.

A private server is required for processing data sets stored locally; since the service
is stateless, providing neither data upload nor storage of reduction
diagrams and reduced data, there is no other way to reduce data that is
not present in the public data repositories.

Local file access is only enabled for servers running on the
private ``localhost'' network interface.
Such servers should not allow external connections since
access to local files is a security risk.

Similarly, a private server is required for custom reduction steps
that are not part of the standard service since the standard
service does not allow the execution of arbitrary Python code.
Users at facilities that do not allow external access to the web will
need to copy all the data files to a local machine and reduce
it with a private server.

\subsubsection{Container-based configuration}
The service can be run using a linux container environment.
A recipe for Docker
is provided 
in the  source code for running the application as three coordinated containers:
one for the web server, one for the Python calculation engine and one for
the Redis cache.  The current snapshot of the source code in the users'
directory is copied into the containers as part of the build step, so this is
a useful setup for development and testing.
The user need only install Docker and Docker-Compose.
The supporting tools, including Python, Redis and all dependent libraries are pulled in by the Docker-Compose recipe.
This greatly eases ability of users to extend the project and to test new features.

\subsubsection{Scalable production server configuration}
For production (public-facing) servers, the static files can be copied from
the \texttt{/static} subfolder of the source repository to a web server
such as Apache or nginx, where requests to the \texttt{/RPC2} path
are forwarded to a pool of Python calculation engines ({\it e.g.,} using uwsgi
to run the calculations with Apache mod\_uwsgi\_proxy acting as the load balancer),
sharing a Redis instance for caching.

An elastic on-demand service could be built from these components as well,
with multiple (replicated) Redis caches and no limit on the number of worker
Python calculation engines that can be employed by a High-Availabilty web
server setup.  The statelessness of the server means that no complicated
synchronization is required between server components.

\section{Web interface}

\begin{figure}
\label{fig:user_interface}
\caption{\emph{reductus} user interface.  Panels (clockwise, from upper left): source data browser, interactive dataflow diagram, and current module parameters; in the center is a plot of the data corresponding to the current module.}
 \includegraphics[width=\linewidth]{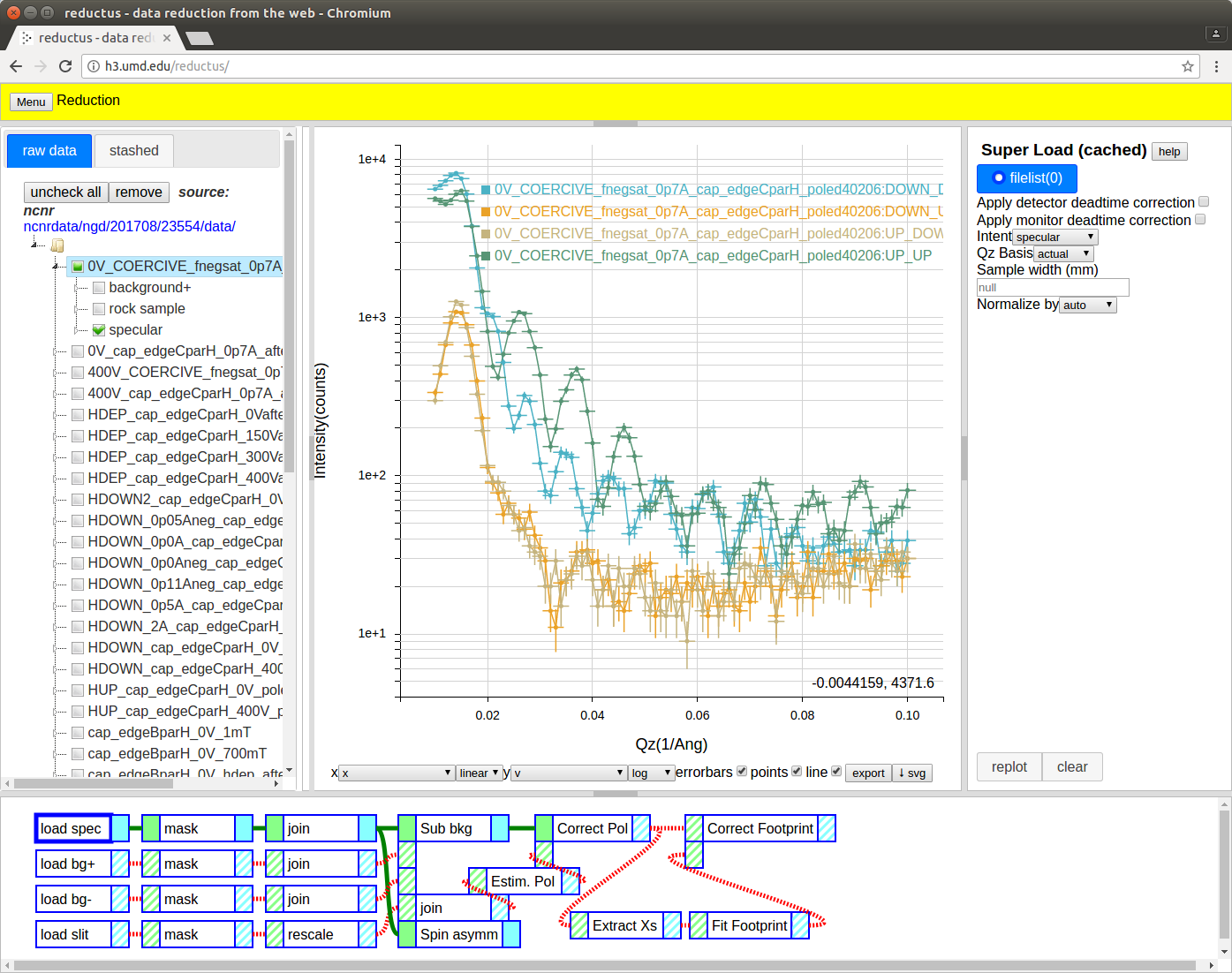}
\end{figure}

The user interface for \emph{reductus} is a javascript application that
runs in a browser. The application relies on a number of advanced
javascript libraries, and as such it is supported only by browsers
that have reasonably standards-compliant implementations of javascript
(\texttt{ECMAScript} version $\geq 5$).
The application is made up of these key components (see Fig. \ref{fig:user_interface}):
\begin{itemize}
\item visual, interactive dataflow diagram
\item panel for setting parameters for individual computation modules
\item plotting panel to show results
\item data source file browser
\end{itemize}

The dataflow display and editor, plotting routines and parameter-setting panels
were implemented using the \texttt{D3.js} visualization library.\cite{bostock2011-d3}

Once connected to the server, the web client can query the computation service for the
public data stores, the available reduction steps, and a set of
predefined template diagrams representing the usual reduction
procedures for the data.  The menus and default options are populated based on
this information.

\subsection{Dataflow diagram}

The user interacts with the dataflow diagram in order to navigate the reduction
chain: by clicking on a module within the diagram to bring up
the parameters panel for that module, or by clicking on the
input (left) or output (right) terminals to display calculation results.
Changes made in the parameters panel are by default immediately applied to the
active template, though by un-checking the \texttt{auto-accept parameters} box
in the \texttt{data} menu an additional confirmation step
(pressing the \fbox{\texttt{accept}} button in the panel)
can be imposed to avoid accidental changing of the active template.
In that context, if a user makes parameter change but then selects a different
module without first pressing \fbox{\texttt{accept}}, that change is lost.

The \fbox{\texttt{clear}} button removes all values from the
active parameters; the client then repopulates the panel with default values
for each parameter.

The web client creates a javascript object notation (JSON) representation of the dataflow diagram
along with an indicator of the input or output of interest,
and sends the request to the server via HTTP POST.
The response is a plottable representation of the data for that connector encoded either as JSON or MSGPACK,
which is then displayed in the plotting panel.

Clicking on any output within the dataflow diagram (including the rightmost, ``final'' output)
will trigger a calculation of
all ancestor results for that result in the diagram on the server, as described in
section \ref{sec:computing_templates}.

So, while the user has the option to inspect intermediate calculation steps, for a routine
reduction dataflow one can simply populate the file inputs and click on the final output for export
the completely reduced data.

\subsection{Parameters panel}

At the beginning of a reduction, a user chooses an instrument definition for working with
the data.  As described in \ref{sec:computing_templates}, this includes a list of
data types and a list of reduction steps (modules) to act on those data types.
The parameters panel is rendered based on the definition for the chosen module type,
using the predefined types for each input field to the module function, mapping the
simple known types (\texttt{int, float, bool,} {\it etc.}) to HTML form elements.
Some field types have renderers with enhanced interaction with the plot panel, such as
an \texttt{index} type which allows clicking on data points to add them to an index list
in the parameters panel.  Another example of enhanced interactions is the
\texttt{scale} type which enables dragging a whole dataset on the
plot to set the scaling factor in the parameters panel.

When parameters are changed in this panel and committed with \fbox{\texttt{accept}}, they
will be used in any calculation of data flowing through that module.

\subsection{Browser caching of calculations}
In addition to the caching provided on the server for avoiding recalculation
of identical results, a local browser cache of calculations is provided.
This is particularly useful for the initial source data load, in which
metadata from all of the files in a source directory are passed to the client for inspection
and sorting in the source file browser.  Naturally in a data reduction scheme,
the amount of data on the input side (loaders) is much greater than the output
result, so caching of the inputs helps tremendously when making small adjustments
interactively to the dataflow algorithm or parameters.

\subsection{Sessions and persistence}

The \emph{reductus} server is stateless; the reduction diagrams created by the user are not stored
(a unique hash of the template representation may be associated with cached calculations on the server,
but no user template is ever stored there).
The only state associated with a session is stored in the browser or on the filesystem of the user's computer.

\subsubsection{Stashing in-browser}

Results of calculations can be ``stashed'' in the local persistent memory of the browser.  A list of these
results can be recalled in the client and reused in two ways: by reloading the entire calculation into the
active dataflow panel, or by selecting multiple stashed results to directly compare their outputs.

\subsubsection{Saving to and loading from filesystem}

\label{subsec:saving}
In addition to the local browser store, the user may download a text version of
the dataflow diagram in JSON format with
\texttt{Menu$\rightarrow$Template$\rightarrow$Download}, which can be reloaded with
\texttt{Menu$\rightarrow$Template$\rightarrow$Upload}.
The file contains the diagram along with any field values, including the names of the input files.
The actual data is not included.

The data for the currently selected node can be saved with \texttt{Menu$\rightarrow$Data$\rightarrow$Export}.
This prompts the user for a filename,
then produces a tab-delimited column-format text file
with the dataflow diagram prepended as a comment header.
The stored diagram allows a full reduction to be reloaded into the
client with \texttt{Menu$\rightarrow$Data$\rightarrow$Reload Exported}
(note that this may trigger recalculation if the raw data has been updated since the reduction was exported).

Due to security limitations built into all current browsers the data may only be saved to the user's ``Downloads''
folder, while uploads can of course originate from any user-readable folder.

\subsubsection{Sharing data among collaborators}

The dataflow diagram is self contained.
The reduced-data text files produced by the \emph{reductus} system can be shared with others easily by email or portable media,
and provide both useful data and a recipe in the header for recreating the data from known sources with known transformations;
the chain of logic can be inspected and verified by reloading the dataflow into the web client at
any time, thus assuring data provenance.
This allows for easy collaboration amongst users without the need for accounts or passwords.



\section{Conclusions}

The \emph{reductus} system is an interesting experiment in providing stateful web services with a stateless server.
Although users lose the convenience of cloud services for managing their data, they are free from the inconvenience of maintaining yet another user ID and password.
Files can be stored and shared using familiar tools such as email. 
The server is easy to adapt and install locally for the rare user that needs more than the rigid set of functions and data sources provided in the remote web service;
this is no more complex than adapting and installing the equivalent desktop application would be.
For the developer, the stateless server needs very little maintenance.
There are no database migrations needed, no backups required, and moving the service to a different computer is as simple as installing the software and redirecting the domain name service (DNS) to an alternate IP.
Javascript provides a flexible environment for interactive applications, with a rich and growing ecosystem of libraries that work across most browser platforms.
The Python backend provides an ecosystem for rapid development of numerical code.
The web services middleware gives us scalability with no additional effort.

Making the dataflow graph visible and modifiable increases flexibility without increasing complexity.
Users with simple reduction problems can enter their data on the left of the graph and retrieve the results on the right, ignoring all steps in between.
If there are problems, they can examine inputs and outputs at each step and identify the cause.
Although 85\% of reduction is performed on-site during the experiment, there were over 150 external users access the system from across the United States and Europe in 2017.
Without the need to install a local version of the software we have far fewer support requests; now they only occur for unusual datasets.

Feedback from the users has been overwhelmingly positive, and the new system has completely supplanted our old reduction software for the
three neutron reflectometry instruments at the NCNR, with work underway to adapt the system to several new instrument classes at the facility.


%
%



\ack{Significant contributions to the early code were made by Brendan Rowan, Alex Yee, Ophir Lifshitz and Elakian Kanakaraj as part of the NIST Summer High school Internship Program (SHIP) and Joseph Redmon as part of the Summer Undergraduate Research Fellowship (SURF), supported by NSF CHRNS grant DMR-1508249.
We are grateful to Prof. Robert Briber of the Dept. of Materials Science and Engineering
at the University of Maryland for providing significant technical support and operating the servers for early deployments of the \emph{reductus} service.

The identification of commercial products or services in this paper does not imply recommendation or endorsement by the National Institute of Standards and Technology, nor does it imply that the equipment or service used is necessarily the best available for the purpose.
}


\bibliographystyle{iucr}
\bibliography{webreduce}





\end{document}